\documentclass[aps,pra,preprint,groupedaddress,showpacs]{revtex4}
\usepackage{bm}
\begin{document}
\preprint{Version 2.0}

\title{Higher-order effective Hamiltonian for light atomic systems}

\author{Krzysztof Pachucki}
\email[]{krp@fuw.edu.pl} \homepage[]{www.fuw.edu.pl/~krp}

\affiliation{Institute of Theoretical Physics, Warsaw University,
             Ho\.{z}a 69, 00-681 Warsaw, Poland}

\date{\today}

\begin{abstract}
We present the derivation of the effective higher-order 
Hamiltonian, which gives $m\,\alpha^6$ contribution 
to energy levels of an arbitrary light atom. The derivation is based
on the Foldy-Wouthuysen transformation of the one-particle Dirac
Hamiltonian followed by  perturbative expansion of the
many particle Green function.
The obtained results can be used for the high precision calculation
of relativistic effects in atomic systems.
\end{abstract}

 \maketitle

\section{Introduction}
The calculation of relativistic corrections to energy levels 
of atomic systems is usually accomplished by using the many-electron
Dirac-Coulomb  (DC) Hamiltonian with possible inclusion of the Breit interaction
between electrons. However, such a Hamiltonian can not be rigorously derived 
from Quantum Electrodynamic (QED) theory, and thus gives an incomplete treatment
of relativistic and QED effects. 
The electron self-energy and vacuum polarization
can be included in the DC Hamiltonian \cite{igd87,iad90}, 
though only in an approximate way. A different approach, which is justified 
by quantum field theory, is to start from a well adapted 
one-electron local potential and build many body perturbation theory. 
This approach allows for the consistent inclusion of QED effects 
as well as a correct treatment of the so called ``negative energy states''. 
It is being pursued by Sapirstein and collaborators \cite{mbptqed}, 
but so far no high accuracy results have been achieved for neutral few
electron atoms. An alternative approach, which is suited for light 
atoms, relies on expansion of energy levels in powers of 
the fine structure constant 
\begin{equation}
E(\alpha) = E^{(2)} + E^{(4)} + E^{(5)} + E^{(6)} + O(\alpha^7), \label{01}
\end{equation}
where $E^{(n)}$ is the contribution of order $m\,\alpha^n$, 
so $E^{(2)}$ is the nonrelativistic energy as given by the Schr\"odinger
Hamiltonian $H^{(2)}\equiv H_0$,  
\begin{equation}
H_0 = \sum_a \biggl(\frac{\vec p_a^{\,2}}{2\,m}-
  \frac{Z\,\alpha}{r_a}\biggr)
  + \sum_{a>b}\sum_b\frac{\alpha}{r_{ab}}\,. \label{02}
\end{equation}
$E^{(4)}$ is the leading relativistic correction
given by the Breit-Pauli Hamiltonian $H^{(4)}$ \cite{bs},
\begin{equation}
E^{(4)} = \langle\phi|H^{(4)}|\phi\rangle.\label{03}
\end{equation}
where
\begin{eqnarray}
H^{(4)} &=&\sum_a \biggl\{\frac{\vec p^4_a}{8\,m^3} +
\frac{ \pi\,Z\,\alpha}{2\,m^2}\,\delta^3(r_a)
+\frac{Z\,\alpha}{4\,m^2}\,
\vec\sigma_a\cdot\frac{\vec r_a}{r_a^3}\times \vec p_a\biggr\}
\nonumber \\
&& +\sum_{a>b}\sum_b \biggl\{
-\frac{ \pi\,\alpha}{m^2}\, \delta^3(r_{ab})
-\frac{\alpha}{2\,m^2}\, p_a^i\,
\biggl(\frac{\delta^{ij}}{r_{ab}}+\frac{r^i_{ab}\,r^j_{ab}}{r^3_{ab}}
\biggr)\, p_b^j \nonumber \\ && -
\frac{2\, \pi\,\alpha}{3\,m^2}\,\vec\sigma_a
\cdot\vec\sigma_b\,\delta^3(r_{ab})
+\frac{\alpha}{4\,m^2}\frac{\sigma_a^i\,\sigma_b^j}
{r_{ab}^3}\,
\biggl(\delta^{ij}-3\,\frac{r_{ab}^i\,r_{ab}^j}{r_{ab}^2}\biggr)
+\frac{\alpha}{4\,m^2\,r_{ab}^3} \nonumber \\ &\times& \biggl[
2\,\bigl(\vec\sigma_a\cdot\vec r_{ab}\times\vec p_b -
\vec\sigma_b\cdot\vec r_{ab}\times\vec p_a\bigr)+
\bigl(\vec\sigma_b\cdot\vec r_{ab}\times\vec p_b -
\vec\sigma_a\cdot\vec
r_{ab}\times\vec p_a\bigr)\biggr]\biggr\}\,.\label{04}
\end{eqnarray}
$E^{(5)}$ is the leading QED correction, which includes Bethe logarithms.
It has been first obtained for hydrogen, for a review see Refs. \cite{sy, eides}.
A few years later $E^{(5)}$ was obtained for the helium atom {\cite{helamb},
see Ref. \cite{simple} for a recent simple rederivation. This result 
can be easily extended 
to arbitrary light atoms, and recently calculations of $E^{(5)}$ have been performed for 
lithium \cite{lithium1, lithium2} and beryllium atoms \cite{berylium}.
$E^{(6)}$ is a higher order relativistic correction and is 
the subject of the present work.
It can be expressed as a sum of three terms,
\begin{equation}
E^{(6)} = \langle\phi|H^{(4)}\,\frac{1}{(E-H_0)'}\,H^{(4)}|\phi\rangle + 
          \langle\phi|H^{(6)}|\phi\rangle +
          \alpha^3 \lambda
          \langle\phi|\sum_{a>b}\sum_b\,\delta^{(3)}(r_{ab})|\phi\rangle,
\label{05}
\end{equation}
where $H^{(6)}$ is an effective Hamiltonian of order $m\,\alpha^6$.
It is well known that the second order correction from the Breit-Pauli Hamiltonian
is divergent since it contains, for example, the Dirac $\delta$ functions.
It is less well known that $H^{(6)}$ also leads to divergent matrix elements,
and yet less well known that in the sum of both terms these divergences almost
cancel out. The additional term containing $\lambda$ is the contribution
coming from the forward scattering three-photon exchange amplitude which cancels
the last divergence in electron-electron interactions, which leads to 
a finite result. The cancellation of divergences requires at first
the inclusion of a regulator, a cut-off in the maximum photon momenta,
which is allowed to go to infinity when all terms are combined together.

The first derivation of $H^{(6)}$ was performed for helium fine-structure
by Douglas and Kroll in \cite{dk}. In this case all matrix elements
were finite because they considered only the splitting of $nP_J$ levels. 
The numerical evaluation of this splitting has been performed 
to a high degree of precision by Yan and Drake in \cite{yandrake}. 
Since calculations of higher order relativistic corrections when singular
matrix elements are present are rather complicated, 
they were first studied in detail for positronium,
the electron-positron system. The $m\,\alpha^6$ contribution
to positronium hyperfine splitting was first obtained 
(without annihilation terms) by Caswell and Lepage in \cite{cl}, where they introduced
a new approach to bound state QED namely, nonrelativistic quantum
electrodynamics (NRQED).
Although their original calculations happened to contain some mistakes
the idea of NRQED was very fruitful, because it simplified enormously
the treatment of bound states.  
Its use has led to significant progress in bound state QED, with the calculation
of the complete three photon-exchange contribution of order $m\,\alpha^6$
to positronium energy levels in \cite{pos1a,pos1b}, and \cite{pos2}. 
It was shown there, that by introducing a regulator, either a photon momentum
cut-off or dimensional regularization, one can derive and calculate
all matrix elements in a consistent way. The agreement between these
calculations and the other purely numerical calculation based on
Bethe-Salpeter equation \cite{pos3,pos4} justify 
the correctness of the effective Hamiltonian or NRQED approaches.
It was quickly found, after the positronium exercise, that a similar
effective Hamiltonian $H^{(6)}$ can be derived for the Helium atom.
Although the derivation of $H^{(6)}$ for S- and P-states of helium 
is rather straightforward \cite{helium1}, the elimination of 
electron-electron singularities
and the calculation of matrix elements is quite involved.
For this reason the first results have been obtained for triplet
states $2^3S_1$ in \cite{slamb} and $2^3P$ \cite{plamb}, where electron-electron
singularities are not present, because the wave function vanishes at $\vec r_1 = \vec
r_2$. Within the dimensional regularization scheme Korobov and Yelkhovsky \cite{ky}
were able to derive a complete set of finite operators and calculate their matrix
elements for the $1^1S_0$ ground state of helium.  
None of these results have been confirmed yet. In this work we present
a simple derivation of effective operators contributing to $H^{(6)}$
for an arbitrary state of arbitrary light atoms. The results obtained agree
for the special cases of the $^1S$, $^3S_1$ and $^3P_J$ levels of helium
with the former result in \cite{dk,helium1}.
Since we do not explicitly eliminate here electron-electron singularities
we were not able to verify the result \cite{ky} for the ground state of helium.

Our derivation consists of three steps. The first step is the
Foldy-Wouthuysen (FW) transformation of a single electron Dirac equation
in an electromagnetic field \cite{itzykson}, performed to the appropriate level of accuracy.
The second step is formal. It is the quantization of the electromagnetic field
interacting with the atom, using the Feynman integration by paths method \cite{itzykson}.
The third step is the derivation of an effective interaction through the
perturbative expansion of the equal-time Green function of the total atomic system.

\section{Foldy-Wouthuysen transformation}
The Foldy-Wouthuysen (FW) transformation \cite{itzykson} is the
nonrelativistic expansion of the Dirac Hamiltonian in an external
electromagnetic field, 
\begin{equation}
H = \vec \alpha \cdot \vec \pi +\beta\,m + e\,A^0\,, \label{06}
\end{equation}
where $\vec \pi = \vec p-e\,\vec A$. The FW
transformation $S$ \cite{itzykson} leads to a new Hamiltonian
\begin{equation}
H_{FW} = e^{i\,S}\,(H-i\,\partial_t)\,e^{-i\,S}\,, \label{07}
\end{equation}
which decouples the upper and lower components of the Dirac wave
function up to a specified order in the $1/m$ expansion. Here 
we calculate FW Hamiltonian up to terms which contribute
to $m\,\alpha^6$ to the energy. While it is not clear here 
which term contributes at which order, we postpone this
to the next section where this issue become more obvious.
Contrary to standard textbooks, we use a more convenient
single Foldy-Wouthuysen operator $S$, which can be written as 
\begin{eqnarray}
S &=&-\frac{i}{2\,m}\,\biggl\{ \beta\,\vec\alpha\cdot\vec\pi -
\frac{1}{3\,m^2}\,\beta\,(\vec\alpha\cdot\vec\pi)^3
+\frac{1}{2\,m}\,[\vec\alpha\cdot\vec\pi\, ,\,
e\,A^0-i\,\partial_t]+Y\biggr\}\,.\label{08}
\end{eqnarray}
where $Y$ is an as yet unspecified odd operator $\{\beta,Y\}=0$, such that
$[Y,e\,A^0-i\,\partial_t]\approx[Y,(\vec\alpha\cdot\vec\pi)^3]\approx 0$. 
It will be fixed at the end to cancel all higher order odd terms.
The F.W. Hamiltonian is expanded in a power series in $S$
\begin{equation}
H_{FW} = \sum_{j=0}^6 {\cal H}^{(j)}+\ldots \label{09}
\end{equation}
where 
\begin{eqnarray}
{\cal H}^{(0)} &=& H, \nonumber \\
{\cal H}^{(1)} &=& [i\,S\,,{\cal H}^{(0)}-i\,\partial_t],\nonumber \\
{\cal H}^{(j)} &=& \frac{1}{j}\,[i\,S\,,{\cal H}^{(j-1)}]\; {\mbox{\rm for $j>1$}},
\label{10}
\end{eqnarray}
and higher order terms in this expansion, denoted by dots, are
neglected. The calculations of subsequent commutators
is rather tedious. For the reader's convenience we present
a separate result for each $H^{(j)}$,
\begin{eqnarray}
{\cal H}^{(1)} &=& \frac{\beta}{m}\,(\vec\alpha\cdot\vec\pi)^2-
            \frac{\beta}{3\,m^3}\,(\vec\alpha\cdot\vec\pi)^4
            -\frac{i\,e}{4\,m^2}[\vec\alpha\cdot\vec\pi,\vec\alpha\cdot\vec E]
            +\frac{1}{2\,m}[Y,\vec\alpha\cdot\vec\pi]\nonumber \\ &&
            -\vec\alpha\cdot\vec\pi + 
            \frac{1}{3\,m^2}\,(\vec\alpha\cdot\vec\pi)^3 - \beta\,Y
            -\frac{\beta}{6\,m^3}[(\vec\alpha\cdot\vec\pi)^3,e\,A^0-i\,\partial_t]
            -\frac{e}{4\,m^2}\,\vec\alpha\cdot\dot{\vec E}\label{11}\\
{\cal H}^{(2)} &=& -\frac{\beta}{2\,m}\,(\vec\alpha\cdot\vec\pi)^2+
            \frac{\beta}{3\,m^3}\,(\vec\alpha\cdot\vec\pi)^4-
            \frac{\beta}{18\,m^5}\,(\vec\alpha\cdot\vec\pi)^6 
            +\frac{i\,e}{8\,m^2}\,[\vec\alpha\cdot\vec\pi,\vec\alpha\cdot\vec
            E]\nonumber \\ &&
            -\frac{1}{2\,m}\,[Y,\vec\alpha\cdot\vec\pi]\nonumber             
            -\frac{i\,e}{24\,m^4}\,[(\vec\alpha\cdot\vec\pi)^3,\vec\alpha\cdot\vec
            E] \nonumber \\ && +\frac{1}{24\,m^4}[\vec\alpha\cdot\vec\pi, 
            [(\vec\alpha\cdot\vec\pi)^3,e\,A^0-i\,\partial_t]]
            -\frac{\beta\,e}{16\,m^3}\,\bigl(\vec\alpha\cdot\vec\pi\,
            \vec\alpha\cdot\dot{\vec E} + \vec\alpha\cdot\dot{\vec E}\,
            \vec\alpha\cdot\vec\pi\bigr)\nonumber \\ &&
            -\frac{1}{2\,m^2}\,(\vec\alpha\cdot\vec\pi)^3+
            \frac{1}{3\,m^4}\,(\vec\alpha\cdot\vec\pi)^5-
            \frac{i\,\beta\,e}{16\,m^3}\,[\vec\alpha\cdot\vec\pi,
            [\vec\alpha\cdot\vec\pi,\vec\alpha\cdot\vec E]]\nonumber \\ &&
            -\frac{i\,\beta\,e}{8\,m^3}\,\bigl((\vec\alpha\cdot\vec\pi)^2\,
            \vec\alpha\cdot\vec E + \vec\alpha\cdot\vec E\,
            (\vec\alpha\cdot\vec\pi)^2\bigr)\label{12}\\
{\cal H}^{(3)} &=& -\frac{\beta}{6\,m^3}\,(\vec\alpha\cdot\vec\pi)^4+
            \frac{\beta}{6\,m^5}\,(\vec\alpha\cdot\vec\pi)^6+
            \frac{i\,e}{96\,m^4}\,[\vec\alpha\cdot\vec\pi,[\vec\alpha\cdot\vec\pi,
            [\vec\alpha\cdot\vec\pi,\vec\alpha\cdot\vec E]]]\nonumber \\ &&
            +\frac{i\,e}{48\,m^4}\,[\vec\alpha\cdot\vec\pi,(\vec\alpha\cdot\vec\pi)^2\,
            \vec\alpha\cdot\vec E + \vec\alpha\cdot\vec
            E\,(\vec\alpha\cdot\vec\pi)^2]
            +\frac{i\,e}{24\,m^4}\,[(\vec\alpha\cdot\vec\pi)^3,
            \vec\alpha\cdot\vec E]\nonumber \\ &&
            +\frac{1}{6\,m^2}\,(\vec\alpha\cdot\vec\pi)^3-\frac{1}{6\,m^4}\,
            (\vec\alpha\cdot\vec\pi)^5+
            \frac{i\,\beta\,e}{48\,m^3}\,[\vec\alpha\cdot\vec\pi,
            [\vec\alpha\cdot\vec\pi,\vec\alpha\cdot\vec E]]\nonumber \\ &&
            +\frac{i\,\beta\,e}{24\,m^3}\,\bigl((\vec\alpha\cdot\vec\pi)^3\,
            \vec\alpha\cdot\vec E + \vec\alpha\cdot\vec
            E\,(\vec\alpha\cdot\vec\pi)^3\bigr)\label{13}\\
{\cal H}^{(4)} &=& \frac{\beta}{24\,m^3}\,(\vec\alpha\cdot\vec\pi)^4-
            \frac{\beta}{18\,m^5}\,(\vec\alpha\cdot\vec\pi)^6-
            \frac{i\,e}{384}\,[\vec\alpha\cdot\vec\pi,[\vec\alpha\cdot\vec\pi,
            [\vec\alpha\cdot\vec\pi,\vec\alpha\cdot\vec E]]]\nonumber \\ &&
            -\frac{i\,e}{192\,m^4}\,[\vec\alpha\cdot\vec\pi,(\vec\alpha\cdot\vec\pi)^2\,
            \vec\alpha\cdot\vec E + \vec\alpha\cdot\vec E\,(\vec\alpha\cdot\vec\pi)^2]-
            \frac{i\,e}{96\,m^4}\,[(\vec\alpha\cdot\vec\pi)^3,\vec\alpha\cdot\vec
            E] \nonumber \\ && +\frac{1}{24\,m^4}\,(\vec\alpha\cdot\vec\pi)^5
            \label{14}\\
{\cal H}^{(5)} &=& -\frac{1}{120\,m^4}\,(\vec\alpha\cdot\vec\pi)^5+
            \frac{\beta}{120\,m^5}\,(\vec\alpha\cdot\vec\pi)^6\label{15}\\
{\cal H}^{(6)} &=& -\frac{\beta}{720\,m^5}\,(\vec\alpha\cdot\vec\pi)^6 \label{16}
\end{eqnarray}
The sum of ${\cal H}^{(i)}$, Eq. (\ref{09}) gives a Hamiltonian, which still
depends on $Y$. Following FW principle, this operator is now chosen to cancel 
all the higher order odd terms from this sum, namely:
\begin{eqnarray}
Y &=&  \frac{\beta}{5\,m^4}\,(\vec\alpha\cdot\vec\pi)^5-
     \frac{\beta\,e}{4\,m^2}\,\vec\alpha\cdot\dot{\vec E}+
     \frac{i\,e}{24\,m^3}\,[\vec\alpha\cdot\vec\pi,
            [\vec\alpha\cdot\vec\pi,\vec\alpha\cdot\vec E]]\nonumber \\ &&
     -\frac{i\,e}{3\,m^3}\,\bigl((\vec\alpha\cdot\vec\pi)^2\,
            \vec\alpha\cdot\vec E + \vec\alpha\cdot\vec
     E\,(\vec\alpha\cdot\vec\pi)^2\bigr). \label{17}
\end{eqnarray}
$Y$ fulfills the initial ansatz, that commutators
$[Y,e\,A^0-i\,\partial_t]$ and $[Y,(\vec\alpha\cdot\vec\pi)^3]$ are of higher
order and thus can be neglected. The resulting FW Hamiltonian is
\begin{eqnarray}
H_{FW} &=& e\,A^0 + \frac{(\vec\sigma\cdot\vec\pi)^2}{2\,m} - 
\frac{(\vec\sigma\cdot\vec\pi)^4}{8\,m^3}
+\frac{(\vec\sigma\cdot\vec\pi)^6}{16\,m^5}
-\frac{i\,e}{8\,m^2}\,[\vec\sigma\cdot\vec\pi,\vec\sigma\cdot\vec E]\nonumber \\ && 
-\frac{e}{16\,m^3}\,\Bigl(\vec\sigma\cdot\vec\pi\,\vec\sigma\cdot\dot{\vec E}+
\vec\sigma\cdot\dot{\vec E}\,\vec\sigma\cdot\vec\pi\Bigr)
-\frac{i\,e}{128\,m^4}\,[\vec\sigma\cdot\vec\pi,[\vec\sigma\cdot\vec\pi,
[\vec\sigma\cdot\vec\pi,\vec\sigma\cdot{\vec E}]]]\nonumber \\ &&
+\frac{i\,e}{16\,m^4}\,\Bigl((\vec\sigma\cdot\vec\pi)^2\,
[\vec\sigma\cdot\vec\pi,\vec\sigma\cdot\vec E] + 
[\vec\sigma\cdot\vec\pi,\vec\sigma\cdot\vec
  E]\,(\vec\sigma\cdot\vec\pi)^2\Bigr),\label{18}
\end{eqnarray}
where we used the commutator identity
\begin{eqnarray}
[(\vec\sigma\cdot\vec\pi)^3,\vec\sigma\cdot\vec E] &=& 
-\frac{1}{2}\,[\vec\sigma\cdot\vec\pi,[\vec\sigma\cdot\vec\pi,
[\vec\sigma\cdot\vec\pi,\vec\sigma\cdot{\vec E}]]]
\nonumber \\ &&
+\frac{3}{2}\,\Bigl((\vec\sigma\cdot\vec\pi)^2\,
[\vec\sigma\cdot\vec\pi,\vec\sigma\cdot\vec E] + 
[\vec\sigma\cdot\vec\pi,\vec\sigma\cdot\vec
  E]\,(\vec\sigma\cdot\vec\pi)^2\Bigr)\label{19}
\end{eqnarray}
to simplify $H_{FW}$. Moreover,
there is some arbitrariness in the operator $S$, what means
that $H_{FW}$ is not unique. The standard approach \cite{itzykson}, which relies
on subsequent use of FW-transformations differs from this one,
by the transformation $S$ with some additional even operator.
However, all $H_{FW}$ have to be equivalent 
at the level of matrix elements between the states which satisfy the
Schr\"odinger equation. 

Let us now study the simple case of an external static potential $V\equiv
e\,A^0$. The FW Hamiltonian with the help of simple commutations takes the form
\begin{eqnarray}
H_{DC} &=& V+\frac{p^2}{2\,m}-\frac{p^4}{8\,m^3}+\frac{p^6}{16\,m^5}
+\frac{1}{8\,m^2}\Bigl(\nabla^2\,V+2\,\vec\nabla V\times\vec
p\cdot\vec\sigma\Bigr)
\nonumber \\ &&
-\frac{3}{32\,m^4}\Bigl(p^2\,\vec\nabla V\times\vec p\cdot\vec\sigma +
                       \vec\nabla V\times\vec p\cdot\vec\sigma\,p^2\Bigr)
+\frac{1}{128\,m^4}\,[p^2,[p^2,V]]\nonumber \\ &&
-\frac{3}{64\,m^4}\,\Bigl(p^2\,\nabla^2 V + \nabla^2 V\, p^2\Bigr).\label{20}
\end{eqnarray}
This Hamiltonian is equivalent to the one
derived previously in \cite{pos1a}, after use of the identity
\begin{equation}
\langle\phi|[p^2,[p^2,V]]|\phi\rangle = 
4\,\langle\phi|(\vec\nabla V)^2|\phi\rangle \label{21}
\end{equation}
which holds for expectation values on stationary Schr\"odinger states $\phi$.
For the exact Coulomb potential $V = -Z\,\alpha/r$, matrix elements
of $H_{DC}$ become singular. Nevertheless, as was  shown in \cite{pos1a},
one can obtain Dirac energy levels up to order $m\,(Z\,\alpha)^6$
by regularizing the Coulomb potential in an arbitrary way, and all
singularities cancel out between the first and second order matrix elements. 

Our aim here is to obtain the Hamiltonian for further calculations
of $m\,\alpha^6$ contribution to energy levels of an arbitrary light atom.
For this one can neglect the vector potential $\vec A$ in all the terms
having $m^4$ and $m^5$ in the denominator. Moreover, less obviously,
one can neglect the term with  $\vec\sigma\cdot\vec A\,\vec\sigma\cdot\dot{\vec E}$ 
and the $\vec B^2$ term. It is because they are of second order in
electromagnetic fields which additionally contain derivatives, 
and thus contribute only at higher orders. 
After these simplifications, $H_{FW}$ takes the form
\begin{eqnarray}
 H_{FW} &=& e\,A^0 + \frac{1}{2\,m}\,\bigl(\pi^2-e\,\vec\sigma\cdot\vec B \bigr) - 
\frac{1}{8\,m^3}\,\bigl(\pi^4-e\,\vec\sigma\cdot\vec B\,\pi^2 -
\pi^2\,e\,\vec\sigma\cdot\vec B\bigr)
\nonumber \\ &&
-\frac{1}{8\,m^2}\Bigl(e\vec\nabla\cdot\vec E + e\,\vec\sigma\cdot
\bigl(\vec E\times\vec\pi-\vec\pi\times\vec E \bigr)\Bigr) 
-\frac{e}{16\,m^3}\,\Bigl(\vec\sigma\cdot\vec p\,\vec\sigma\cdot\dot{\vec E}+
\vec\sigma\cdot\dot{\vec E}\,\vec\sigma\cdot\vec p\Bigr)
\nonumber \\ &&
-\frac{3}{32\,m^4}\Bigl(p^2\,\vec\nabla (e\,A^0)\times\vec p\cdot\vec\sigma +
                       \vec\nabla (e\,A^0)\times\vec p\cdot\vec\sigma\,p^2\Bigr)
+\frac{1}{128\,m^4}\,[p^2,[p^2,e\,A^0]]\nonumber \\ &&
-\frac{3}{64\,m^4}\,\Bigl(p^2\,\nabla^2 (e\,A^0) + \nabla^2 (e\,A^0)\,
p^2\Bigr) +\frac{1}{16\,m^5}\,p^6 \label{22}
\end{eqnarray}
From this Hamiltonian one builds the many body Lagrangian density
\begin{equation}
{\cal L} = \phi^\star(i\,\partial_t-H_{FW})\phi + {\cal L}_{EM},\label{23}
\end{equation}
where ${\cal L}_{EM}$ is a Lagrangian of the electromagnetic field,
and with the help of perturbation theory  calculates Green functions.
 
\section{The higher order Breit-Pauli Hamiltonian}
We consider the equal time retarded Green function $G=G(\{\vec r'_a\}, t'; 
\{\vec r_a\},t)$, where by $ \{\vec r_a\}$ we denote the set of coordinates
for all particles of the system. This Green function is similar to that
used by Shabaev in \cite{shabaev}. In the stationary case considered here,
$G = G(t'-t)$. The Fourier transform of $G$ in the time variable $t'-t$
can be written as
\begin{equation}
G(E) \equiv \frac{1}{E-H_{\rm eff}(E)}\label{24}
\end{equation}
which is the definition of the effective Hamiltonian $H_{\rm eff}(E)$.
In the nonrelativistic case $H_{\rm eff}=H_0$.
All the relativistic and QED corrections resulting from the Lagrangian
can be represented as
\begin{eqnarray}
G(E) &=& \frac{1}{E-H_0}+\frac{1}{E-H_0}\,\Sigma(E)\,\frac{1}{E-H_0} +
\frac{1}{E-H_0}\,\Sigma(E)\,\frac{1}{E-H_0}\,\Sigma(E)\,\frac{1}{E-H_0}+\ldots
\nonumber \\
&=& \frac{1}{E-H_0-\Sigma(E)} \equiv \frac{1}{E-H_{\rm eff}(E)}\label{25}
\end{eqnarray}
where $\Sigma(E)$ is the $n$-particle irreducible contribution.
The energy level can be interpreted as a pole of $G(E)$
as a function of $E$. For this it is convenient to 
consider the matrix element of $G$ between the
the nonrelativistic wave function corresponding to this 
energy level. There is always such a correspondence, since
relativistic and QED effects are small perturbations of the system. 
We follow here a relativistic  approach for the electron self-energy presented
in \cite{sy}. This matrix element is
\begin{equation}
\langle\phi|G(E)|\phi\rangle = 
\langle\phi|\frac{1}{E-H_0-\Sigma(E)}|\phi\rangle \equiv 
\frac{1}{E-E_0-\sigma(E)}\label{26}
\end{equation}
where 
\begin{equation}
\sigma(E) = \langle\phi|\Sigma(E)|\phi\rangle + 
\sum_{n\neq 0}\langle\phi|\Sigma(E)|\phi_n\rangle
\,\frac{1}{E-E_n}\,\langle\phi_n|\Sigma(E)|\phi\rangle
+\ldots\label{27}
\end{equation}
Having $\sigma(E)$, the correction to the energy level 
can be expressed as
\begin{eqnarray}
\delta E &=& E-E_0 = \sigma(E_0) +\sigma'(E_0)\, \sigma(E_0)+\ldots
\nonumber \\ &=&
\langle\phi|\Sigma(E_0)|\phi\rangle + 
\langle\phi|\Sigma(E_0)\,\frac{1}{(E_0-H_0)'}\,\Sigma(E_0)|\phi\rangle+
\langle\phi|\Sigma'(E_0)|\phi\rangle\,\langle\phi|\Sigma(E_0)|\phi\rangle 
+\ldots\label{28}
\end{eqnarray}
Since the last term in Eq. (\ref{28}) can be neglected up to order $m\,\alpha^6$, 
one can consider only $\Sigma(E_0)$. In most cases, the explicit dependence 
of $\Sigma$ on state, through $E_0$, can be eliminated by appropriate
transformations, with the help of various commutations. 
The only exception is the so called Bethe logarithm, which contributes
only to the order $m\,\alpha^5$. If we consider this term separately,
the operator $\Sigma$ gives an effective Hamiltonian
\begin{equation}
H_{\rm eff} = H_0 + \Sigma  = H_0 + H^{(4)} + H^{(5)} +  H^{(6)} + \ldots 
\label{29}
\end{equation}
from which one calculates corrections to energy levels as in Eq. (\ref{05}).
The calculation of $\Sigma$ follows from Feynman rules for
Lagrangian in Eq. (\ref{23}). We will use the photon propagator in the Coulomb gauge:
\begin{eqnarray}
G_{\mu\nu}(k) = \left\{
\begin{array}{ll}
{-\frac{1}{\vec{k}^2}} & \mu = \nu = 0\,, \\
{\frac{-1}{k_0^2-\vec k^2 +i\,\epsilon}}\Bigl(\delta_{ij}-{\frac{{k}_i {k}_j}
{\vec{k}^2}\Bigr)} & \mu =i, \nu =j
\end{array}
\right.\,. \label{30}
\end{eqnarray}
and consider separately corrections due to exchange of the Coulomb $G_{00}$
and the transverse $G_{ij}$ photon. The typical one photon exchange
contribution between electrons $a$ and $b$ is:
\begin{eqnarray}
\langle\phi|\Sigma(E_0)|\phi\rangle &=& e^2\int\frac{d^4
k}{(2\,\pi)^4\,i}\,G_{\mu\nu}(k)\,\biggl\{
\biggl\langle\phi\biggl|\jmath^\mu_a(k)\,e^{i\,\vec k\cdot\vec
r_a} \,\frac{1}{E_0-H_0-k^0+i\,\epsilon}
\,\jmath^\nu_b(-k)\,e^{-i\,\vec k\cdot\vec
r_b}\,\biggr|\phi\biggr\rangle \nonumber \\ &&
+\biggl\langle\phi\biggl|\jmath^\mu_b(k)\,e^{i\,\vec k\cdot\vec
r_b} \,\frac{1}{E_0-H_0-k^0+i\,\epsilon}
\,\jmath^\nu_a(-k)\,e^{-i\,\vec k\cdot\vec
r_a}\,\biggr|\phi\biggr\rangle \biggr\}\,,\label{31}
\end{eqnarray}
where $\phi$ is an eigenstate of $H_0$ and $\jmath^\mu_a$ is an
electromagnetic current operator for particle $a$. One obtains
the exact form of  $\jmath^\mu(k)$ from the Lagrangian in Eq.(\ref{23}),
and is defined as the coefficient which multiplies
the polarization vector $\epsilon^\mu$ in the annihilation part
of the electromagnetic potential
\begin{equation}
A^\mu(\vec r,t) \sim \epsilon^\mu_\lambda\, e^{i\,\vec k\cdot\vec r -
  i\,k^0\,t}\,
\hat a_\lambda +{\rm h.c.}\label{32}
\end{equation}
The first terms of the nonrelativistic expansion of $\jmath^0$ component are
\begin{equation}
\jmath^0(\vec k) = 1 +\frac{i}{4\,m}\,\vec\sigma\cdot\vec
k\times\vec p - \frac{1}{8\,m^2}\vec k^{\,2}+\ldots\label{33}
\end{equation}
and of the $\vec\jmath$ component are
\begin{equation}
\vec \jmath(\vec k) = \frac{\vec p}{m} +
\frac{i}{2\,m}\,\vec\sigma\times\vec k\,.\label{34}
\end{equation}
Most of the calculation is performed in the nonretardation approximation,
namely one sets $k^0=0$ in the photon propagator $G_{\mu\nu}(k)$ and $\jmath(k)$. 
The retardation corrections are considered separately.
Within this approximation and using the symmetrization $k^0 \leftrightarrow -k^0$, 
the $k^0$ integral is
\begin{equation}
\frac{1}{2} \int\frac{d\,k^0}{2\,\pi\,i}\,
\biggl[\frac{1}{-\Delta E-k^0+i\,\epsilon}+
\frac{1}{-\Delta E+k^0+i\,\epsilon}\biggr] = -\frac{1}{2}\label{35}
\end{equation}
where we have assumed that $\Delta E$ is positive,  which is correct
when $\phi$ is the ground state. For excited states, the integration
contour is deformed in such a way that all the poles from the
electron propagator lie on one side, so it is not strictly speaking
the Feynman contour. However the result of the $k^0$ integration
for excited states is the same as in the above, which leads to
\begin{equation}
\langle\phi|\Sigma(E_0)|\phi\rangle = -e^2\int\frac{d^3
k}{(2\,\pi)^3}\,G_{\mu\nu}(\vec k)\,
\biggl\langle\phi\biggl|\jmath^\mu_a(\vec k)\,e^{i\,\vec
k\cdot(\vec r_a-\vec r_b)} \,\jmath^\nu_b(-\vec
k)\,\biggr|\phi\biggr\rangle\,.\label{36}
\end{equation}
The $\vec k$ integral is the Fourier transform
of the photon propagator in the nonretardation approximation
\begin{eqnarray}
G_{\mu\nu}(\vec r) = \int \frac{d^3 k}{(2\,\pi)^3} G_{\mu\nu}(\vec
k) = \frac{1}{4\,\pi}\,\left\{
\begin{array}{ll}
-\frac{1}{r}  & \mu = \nu = 0\,, \\
\frac{1}{2\,r}\Bigl(\delta_{ij}+{\frac{{r}_i {r}_j}
{\vec{r}^{\,2}}\Bigr)}& \mu =i, \nu =j
 \end{array}
\right.\,.\label{37}
\end{eqnarray}
One easily recognizes that in the nonrelativistic limit
$G_{00}$ is the Coulomb interaction. However this term is already included
in $H_0$, which means that this nonrelativistic Coulomb interaction
has to be excluded from the perturbative expansion. Next order terms
resulting from $\jmath^0$ and $\vec\jmath$ lead to the Breit
Pauli Hamiltonian, Eq. (\ref{04}). Below we derive the higher order
term in the nonrelativistic expansion, namely
the $m\,\alpha^6$ Hamiltonian, which we call here the
higher order effective Hamiltonian $H^{(6)}$. 
It is expressed as a sum of various contributions 
\begin{equation}
H^{(6)} =\sum_{i=0,9} \delta H_i\label{38}
\end{equation}
which are calculated in the following.

$\delta H_{0}$ is the  kinetic energy correction
\begin{equation}
\delta H_{0} = \sum_a\frac{p_a^6}{16\,m^5}\,.\label{39}
\end{equation}

$\delta H_{1}$ is a correction due to the last three terms
in $H_{FW}$ in Eq. (\ref{22}). These terms involve only $A^0$, so the nonretardation approximation 
is strictly valid here. This correction $\delta H_{1}$ includes the Coulomb interaction
between the electron and the nucleus, and between electrons.
So, if we denote by  $V$ the nonrelativistic interaction potential 
\begin{equation}
V \equiv \sum_a -\frac{Z\,\alpha}{r_a} + \sum_{a>b}\,\sum_b\frac{\alpha}{r_{ab}}
\label{40}
\end{equation}
and by ${\cal E}_a$ the static electric field at the position of particle $a$
\begin{equation}
e\,\vec{\cal E}_a \equiv -\nabla_a V =  
-Z\,\alpha\,\frac{\vec r_a}{r_a^3} +\sum_{b\neq a}\alpha\,\frac{\vec r_{ab}}{r_{ab}^3}
\label{41}
\end{equation}
then $\delta H_{1}$ can be written as
\begin{eqnarray}
\delta H_{1} &=& \sum_a
\frac{3}{32\,m^4}\,\vec\sigma_a\cdot
\Bigl(p_a^2\,e\,\vec{\cal E}_a\times\vec p_a +
                       e\,\vec{\cal E}_a\times\vec p_a\,p_a^2\Bigr)
\nonumber \\ &&
+\frac{1}{128\,m^4}\,[p_a^2,[p_a^2,V]]
-\frac{3}{64\,m^4}\,\Bigl(p_a^2\,\nabla_a^2 V + \nabla_a^2 V\, p_a^2\Bigr)
\label{42}
\end{eqnarray}

$\delta H_{2}$ is a correction 
to the Coulomb interaction between electrons
which comes from the 4$^{\rm th}$ term in $H_{FW}$, namely
\begin{equation}
-\frac{1}{8\,m^2}\Bigl(e\vec\nabla\cdot\vec E + e\,\vec\sigma\cdot
\bigl(\vec E\times\vec p-\vec p\times\vec E \bigr)\Bigr)
\label{43}
\end{equation}
If interaction of both electrons is modified by this term,
it can be obtained in the nonretardation approximation Eq. (\ref{36}),
so one obtains
\begin{eqnarray}
\delta H_{2} &=& \sum_{a>b}\sum_b
\int d^3 k\,\frac{4\,\pi}{k^2}\,\frac{1}{64\,m^4}\,
\biggl(k^2 +2\,i\,\vec\sigma_a\cdot\vec p_a\times\vec k\biggr)\,
e^{i\,\vec k\cdot\vec r_{ab}}\,
\biggl(k^2 +2\,i\,\vec\sigma_b\cdot\vec k\times\vec p_b\biggr)
\nonumber \\ &=& \sum_{a>b}\sum_b
\frac{1}{64\,m^4}\,\biggl\{
-4\,\pi\,\nabla^2\,\delta^3(r_{ab})
-8\,\pi\,i\,\vec\sigma_a\cdot\vec p_a\times\delta^3(r_{ab})\,\vec p_a
-8\,\pi\,i\,\vec\sigma_b\cdot\vec p_b\times\delta^3(r_{ab})\,\vec p_b
\nonumber \\ &&
+4\,(\vec\sigma_a\times\vec p_a)^i
\biggl[ \frac{\delta^{ij}}{3}\,4\,\pi\,\delta^3(r_{ab})+
\frac{1}{r_{ab}^3}\,\biggl(\delta^{ij}-3\,\frac{r_{ab}^i\,r_{ab}^j}{r_{ab}^2}
\biggr)\biggr] (\vec\sigma_b\times\vec p_b)^j
\biggr\}\label{44}
\end{eqnarray}
We have encountered here for the first time singular electron-electron
operators. One can make them meaningful by appropriate
regularization of the photon propagator, or by dimensional regularization. 
In general it is a difficult problem and,  as we have written in the
introduction, the explicit solution was demonstrated 
for the positronium and helium atom only. 

$\delta H_{3}$ is an correction that comes from 5$^{\rm th}$ term in Eq. (\ref{22}), 
\begin{equation}
-\frac{e}{16\,m^3}\,\Bigl(\vec\sigma\cdot\vec p\,\vec\sigma\cdot\dot{\vec E}+
\vec\sigma\cdot\dot{\vec E}\,\vec\sigma\cdot\vec p\Bigr)\,.\label{45}
\end{equation}
To calculate it, we have to return 
to the original expression for one-photon exchange. We assume
that particle $a$ interacts by this term, while particle $b$ by 
nonrelativistic coupling $e\,A^0$ and obtain
\begin{eqnarray}   
\delta E_{3} &=& \sum_{a\neq b}\sum_b\,
-e^2\int\frac{d^4k}{(2\,\pi)^4\,i}\,
\frac{1}{\vec k^2}\,\frac{1}{16\,m^3}
\nonumber \\ &&\biggl\{\langle\phi|\Bigl(
\vec\sigma_a\cdot\vec p_a\,\vec\sigma_a\cdot\vec k\,
e^{i\,\vec k\cdot \vec r_a} + e^{i\,\vec k\cdot \vec r_a}\,
\vec\sigma_a\cdot\vec p_a\,\vec\sigma_a\cdot\vec k\Bigr)\,
\frac{k^0}{E_0-H_0-k^0+i\,\epsilon}\,e^{-i\,\vec k\cdot \vec r_b}|\phi\rangle
\nonumber \\ &&
+\langle\phi|e^{i\,\vec k\cdot \vec r_b}\,
\frac{k^0}{E_0-H_0-k^0+i\,\epsilon}\,\Bigl(
\vec\sigma_a\cdot\vec p_a\,\vec\sigma_a\cdot\vec k\,
e^{i\,\vec k\cdot \vec r_a} + e^{i\,\vec k\cdot \vec r_a}\,
\vec\sigma_a\cdot\vec p_a\,\vec\sigma_a\cdot\vec k\Bigr)
|\phi\rangle\biggr\}\label{46}
\end{eqnarray}
We replace $\vec k \rightarrow -\vec k$ in the second
term, then perform the $k^0$ integral, and obtain
\begin{eqnarray}   
\delta E_{3} &=& \sum_{a\neq b}\sum_b\,
-\frac{e^2}{2}\,\int\frac{d^3k}{(2\,\pi)^3}\,
\frac{1}{\vec k^2}\,\frac{1}{16\,m^3}
\nonumber \\ &&\biggl\{\langle\phi|\Bigl(
\vec\sigma_a\cdot\vec p_a\,\vec\sigma_a\cdot\vec k\,
e^{i\,\vec k\cdot \vec r_a} + e^{i\,\vec k\cdot \vec r_a}\,
\vec\sigma_a\cdot\vec p_a\,\vec\sigma_a\cdot\vec k\Bigr)\,
(H_0-E_0)\,e^{-i\,\vec k\cdot \vec r_b}|\phi\rangle
\nonumber \\ &&
+\langle\phi|e^{i\,\vec k\cdot \vec r_b}\,
(H_0-E_0)\,\Bigl(
\vec\sigma_a\cdot\vec p_a\,\vec\sigma_a\cdot\vec k\,
e^{i\,\vec k\cdot \vec r_a} + e^{i\,\vec k\cdot \vec r_a}\,
\vec\sigma_a\cdot\vec p_a\,\vec\sigma_a\cdot\vec k\Bigr)
|\phi\rangle\biggr\}\label{47}
\end{eqnarray}
After commuting $(H_0-E_0)$ with $e^{\pm i\,\vec k\cdot \vec r_b}$
one expresses this correction in terms of an effective operator
\begin{equation}
\delta H_3 = \sum_{a\neq b}\sum_b -\frac{1}{32\,m^4}\,
\biggl[p_b^2,\biggl[p_a^2,\frac{\alpha}{r_{ab}}\biggr]\biggr]
\label{48}
\end{equation}

$\delta H_{4}$ is the relativistic correction 
to transverse photon exchange. 
The first electron is coupled to $\vec A$ by the nonrelativistic term
\begin{equation}
-\frac{e}{m}\,\vec p\cdot\vec A -\frac{e}{2\,m}\,\vec\sigma\cdot\vec B
\label{49}
\end{equation} 
and the second one by the relativistic correction, the 3$^{\rm rd}$ term in Eq. (\ref{22})
\begin{equation}
-\frac{1}{8\,m^3}\,\bigl(\pi^4-e\,\vec\sigma\cdot\vec B\,\pi^2
 -\pi^2\,e\,\vec\sigma\cdot\vec B\bigr)
\rightarrow
\frac{e}{8\,m^3}\,\bigl(p^2\,2\,\vec p\cdot\vec A+2\,\vec p\cdot\vec A\,p^2+ 
\vec\sigma\cdot\vec B\,p^2+p^2\,\vec\sigma\cdot\vec B\bigr)\label{50}
\end{equation}
It is sufficient to calculate it in the nonretardation approximation
\begin{eqnarray}
\delta H_4 &=& \sum_{a\neq b}\,\sum_b\,\frac{\alpha}{8\,m^3}\,
\Bigl[2\,p_a^2\,p_a^i + p_a^2\,(\vec\sigma_a\times\nabla_a)^i\Bigr]
\nonumber \\ &&
\Bigl[\frac{p_b^j}{m} + \frac{1}{2\,m}\,(\vec\sigma_b\times\,\nabla_b)^j\Bigr]\,
\frac{1}{2\,r_{ab}}\biggl(\delta^{ij}+\frac{r_{ab}^i\,r_{ab}^j}{r_{ab}^2}\biggr)
+ {\rm h.c.} \label{51}
\end{eqnarray}
It is convenient at this point to introduce a notation
for the vector potential at the position of particle  $a$ 
which is produced by other particles
\begin{equation}
e\,{\cal A}^i_{a} \equiv \sum_{b\neq a} \frac{\alpha}{2\,r_{ab}}
\biggl(\delta^{ij}+\frac{r_{ab}^i\,r_{ab}^j}{r_{ab}^2}\biggr)\,
\frac{p_b^j}{m} + \frac{\alpha}{2\,m}\frac{\bigl(\vec\sigma_b\times\vec
  r_{ab}\bigr)^i}{r_{ab}^3}\,,\label{52}
\end{equation}
then this correction can be written as
\begin{eqnarray}
\delta H_4 &=&
\sum_a\,\frac{e}{8\,m^3}\,
\Bigl[2\,p_a^2\,\vec p_a\cdot\vec {\cal A}_{a} 
+2\,p_a\cdot\vec {\cal A}_{a}\,p_a^2
+ p_a^2\,\vec\sigma_a\cdot\nabla_a\times\vec {\cal A}_{a} 
+ \vec\sigma_a\cdot\nabla_a\times\vec {\cal A}_{a}\,p_a^2\Bigr]\label{53}
\end{eqnarray}
Let us notice that in the nonretardation approximation  any correction
can be simply obtained by replacing the magnetic field $\vec A$ by 
a static field $\vec {\cal A}_{a}$. We will use this fact in further calculations.

$\delta H_{5}$ comes from the coupling
\begin{equation}
\frac{e^2}{8\,m^2}\,\vec\sigma\cdot(\vec E\times\vec A-\vec A\times\vec E)
\label{54}
\end{equation}
 which is present in 4$^{\rm th}$ term in Eq. (\ref{22}). 
The resulting correction is obtained by replacing the fields $\vec E$ and
$\vec A$ by the static fields produced by other electrons
\begin{equation}
\delta H_5 = \sum_a \frac{e^2}{8\,m^2}\,\vec\sigma_a\cdot
\Bigl[\vec{\cal E}_a\times \vec {\cal A}_{a} - \vec {\cal
    A}_{a}\times\vec{\cal E}_a\Bigr]\label{55}
\end{equation}

$\delta H_{6}$ comes from the coupling
\begin{equation}
\frac{e^2}{2\,m}\,\vec A^2\label{56}
\end{equation}
which is present in the second term of Eq. (\ref{22}). Again, in the nonretardation
approximation the $\vec A_a$ field is being replaced by the static fields produced
by other electrons
\begin{equation}
\delta H_6 = \sum_a\frac{e^2}{2\,m^2}\,\vec {\cal A}_a^2 \label{57}
\end{equation}

$\delta H_{7}$ is a retardation correction in the nonrelativistic single
transverse photon exchange. To calculate it, we have to return to the 
general one-photon exchange expression, Eq. (\ref{31}), and take the transverse
part of the photon propagator
\begin{eqnarray}
\delta E &=& -e^2\,\int\frac{d^4k}{(2\,\pi)^4\,i}\,
\frac{1}{(k^0)^2-\vec k^2+i\,\epsilon}\,\biggl(\delta^{ij}-\frac{k^i\,k^j}
{\vec k^2}\biggr)\,
\nonumber \\ &&
\biggl\langle\phi\biggl|\jmath^i_a(k)\,e^{i\,\vec k\cdot\vec
r_a} \,\frac{1}{E_0-H_0-k^0+i\,\epsilon}
\,\jmath^j_b(-k)\,e^{-i\,\vec k\cdot\vec
r_b}\,\biggr|\phi\biggr\rangle +(a\leftrightarrow b)\,.\label{58}
\end{eqnarray}
We assume that the product $ \jmath^i_a(k)\;
\jmath^j_b(-k)$ contains at most a single power of $k^0$.
This allows one to perform the $k^0$ integration by encircling  
the only pole $k^0 = |\vec k|$ on $\Re(k^0)>0$ complex half plane
and obtain
\begin{eqnarray}
\delta E &=& e^2\,\int\frac{d^3k}{(2\,\pi)^3\,2\,k}\,
\biggl(\delta^{ij}-\frac{k^i\,k^j}{k^2}\biggr)\,
\nonumber \\ &&
\biggl\langle\phi\biggl|\jmath^i_a(k)\,e^{i\,\vec k\cdot\vec r_a} \,
\frac{1}{E_0-H_0-k}
\,\jmath^j_b(-k)\,e^{-i\,\vec k\cdot\vec
r_b}\,\biggr|\phi\biggr\rangle +(a\leftrightarrow b)\,.\label{59}
\end{eqnarray}
where $k = |\vec k|$. By using the nonrelativistic
form of $\jmath^i$ and taking the third term in the
retardation expansion, 
\begin{equation} 
\frac{1}{E_0-H_0-k} = -\frac{1}{k}+\frac{H_0-E_0}{k^2}
-\frac{(H_0-E_0)^2}{k^3}+\ldots\label{60}
\end{equation}
where the first one contributes to the Breit-Pauli Hamiltonian, 
the second term to $E^{(5)}$, and the third term gives $\delta E_{7}$
\begin{eqnarray}
\delta E_7 &=& \sum_{a \neq b}\sum_b -e^2\,\int\frac{d^3k}{(2\,\pi)^3\,2\,k^4}\,
\biggl(\delta^{ij}-\frac{k^i\,k^j}{k^2}\biggr)\,
\biggl\langle\phi\biggl|
\biggl(\frac{\vec p_a}{m}+\frac{1}{2\,m}\,
\vec \sigma_a\times\nabla_a\biggr)^i
\,e^{i\,\vec k\cdot\vec r_a}\,
\nonumber \\ &&
(H_0-E_0)^2
\,\biggl(\frac{\vec p_b}{m}+\frac{1}{2\,m}\,
\vec \sigma_b\times\nabla_b\biggr)^j
\,e^{-i\,\vec k\cdot\vec
r_b}\,\biggr|\phi\biggr\rangle\,.\label{61}
\end{eqnarray}
This is the most complicated term in the evaluation,
and we have to split it into three parts with no spin, single spin
and double spin terms
\begin{equation}
\delta E_7 = \delta E_{A}+\delta E_{B}+\delta E_{C}\label{62}
\end{equation}
The part with double spin operators is
\begin{eqnarray}
\delta E_C &=& \sum_a \sum_{a\neq b}-e^2\int\frac{d^3k}{(2\,\pi)^3\,2\,k^4}
\frac{(\vec\sigma_a\times\vec k)\cdot(\vec\sigma_b\times\vec k)}{4\,m^2}\,
\Bigl\langle\phi\Bigl|
\,e^{i\,\vec k\cdot\vec r_a}\,(H_0-E_0)^2
\,e^{-i\,\vec k\cdot\vec
r_b}\,\Bigr|\phi\Bigr\rangle\label{63}
\end{eqnarray}
One uses the commutation identity
\begin{eqnarray}
\Bigl\langle\,e^{i\,\vec k\cdot\vec r_a}\,(H_0-E_0)^2\,e^{-i\,\vec k\cdot\vec r_b}\Bigr\rangle 
+(a\leftrightarrow b)&=& 
\Bigl\langle\Bigl[e^{i\,\vec k\cdot\vec r_a},\Bigl[(H_0-E_0)^2, e^{-i\,\vec
      k\cdot\vec r_b}\Bigr]\Bigr]\Bigr\rangle
\nonumber \\ &=& 
-\frac{1}{2\,m^2}\,\Bigl\langle
\bigl[p_a^2,\bigl[p_b^2,e^{i\,\vec k\cdot\vec r_{ab}}\bigr]\bigr]
\Bigr\rangle\label{64}
\end{eqnarray}
to express this correction in terms of the effective operator $\delta H_{C}$.
\begin{eqnarray}
\delta H_{C} &=& \sum_{a>b}\sum_b\frac{\alpha}{16\,m^4}\,
\biggl[p_a^2,\biggl[p_b^2,
    \vec\sigma_a\cdot\vec\sigma_b\,\frac{2}{3\,r_{ab}}+
\sigma_a^i\,\sigma_b^j\,\frac{1}{2\,r_{ab}}\biggl(\frac{r_{ab}^i\,r_{ab}^j}{r_{ab}^2}
-\frac{\delta^{ij}}{3}\biggr)\biggr]\biggr]\label{65}
\end{eqnarray}
The part with no spin operator is
\begin{eqnarray}
\delta E_A &=& \sum_{a\neq b}\sum_b -e^2\int\frac{d^3k}{(2\,\pi)^3\,2\,k^4}\,
\biggl(\delta^{ij}-\frac{k^i\,k^j}{k^2}\biggr)\,
\nonumber \\ &&
\biggl\langle\phi\biggl|\frac{p_a^i}{m}
\Bigl\{e^{i\,\vec k\cdot\vec r_a}\,(H_0-E_0)^2\,e^{-i\,\vec k\cdot\vec r_b}
-(H_0-E_0)^2\Bigr\}\,\frac{p_b^j}{m}\,
\biggr|\phi\biggr\rangle\,.\label{66}
\end{eqnarray}
We subtracted here the term with $k=0$. We ought to perform this
in Eq. (\ref{61}), where lower order terms were subtracted, but for simplicity
of writing we have not done it until now. We use another commutator identity
\begin{eqnarray}
&&e^{i\,\vec k\cdot\vec r_a}\,(H_0-E_0)^2\,e^{-i\,\vec k\cdot\vec r_b}
-(H_0-E_0)^2 =
\nonumber \\ &&
(H_0-E_0)\,(e^{i\,\vec k\cdot\vec r_{ab}}-1)\,(H_0-E_0) 
+ (H_0-E_0)\,\biggl[\frac{p_b^2}{2\,m},e^{i\,\vec k\cdot\vec r_{ab}}-1\biggr]
\nonumber \\ && + \biggl[e^{i\,\vec k\cdot\vec r_{ab}}-1,\frac{p_a^2}{2\,m}\biggr]\,(H_0-E_0)
+ \biggl[\frac{p_b^2}{2\,m},\biggl[e^{i\,\vec k\cdot\vec
      r_{ab}}-1,\frac{p_a^2}{2\,m}\biggr]\biggr] \label{67}
\end{eqnarray}
and the integration formula
\begin{equation}
\int d^3k\frac{4\,\pi}{k^4}\,\biggl(\delta^{ij}-\frac{k^i\,k^j}{k^2}\biggr)\,
\bigl(e^{i\,\vec k\cdot\vec r}-1\bigr) = \frac{1}{8\,r}\,
\bigl(r^i\,r^j-3\,\delta^{ij}\,r^2\bigr)\label{68}
\end{equation}
to obtain the effective operator $\delta H_A$
\begin{eqnarray}
\delta H_A &=& \sum_{a>b}\sum_b -\frac{\alpha}{8\,m^2}\,\biggl\{
\bigl[p_a^i,V\bigr]\,\frac{r_{ab}^i\,r_{ab}^j-3\,\delta^{ij}\,r_{ab}^2}{r_{ab}}\,
\bigl[V,p_b^j\bigr]\nonumber \\ &&
+\bigl[p_a^i,V\bigr]\,\biggl[\frac{p_b^2}{2\,m},
 \frac{r_{ab}^i\,r_{ab}^j-3\,\delta^{ij}\,r_{ab}^2}{r_{ab}}\biggr]\,p_b^j
+p_a^i\,\biggl[\frac{r_{ab}^i\,r_{ab}^j-3\,\delta^{ij}\,r_{ab}^2}{r_{ab}},
\frac{p_a^2}{2\,m}\biggr]\,\bigl[V,p_b^j\bigr]
\nonumber \\ &&
+p_a^i\,\biggl[\frac{p_b^2}{2\,m},\biggl[
\frac{r_{ab}^i\,r_{ab}^j-3\,\delta^{ij}\,r_{ab}^2}{r_{ab}},
\frac{p_a^2}{2\,m}\biggr]\biggr]\, p_b^j
\biggr\}\label{69}
\end{eqnarray}
The part with the single spin operator is
\begin{eqnarray}
\delta E_B &=& \sum_{a\neq b}\sum_b-\frac{i\,e^2}{4\,m^2}
\int\frac{d^3 k}{(2\,\pi)^3\,k^4}
\nonumber \\ &&
\Bigl\{e^{i\,\vec k\cdot\vec r_a}\,
(H_0-E_0)^2\,e^{-i\,\vec k\cdot\vec r_b}\,\vec\sigma_a\times\vec k\cdot\vec p_b
-\vec p_a\cdot\vec\sigma_b\times\vec k\,e^{i\,\vec k\cdot\vec r_a}\,
(H_0-E_0)^2\,e^{-i\,\vec k\cdot\vec r_b}\Bigr\}\label{70}
\end{eqnarray}
With the help of the commutator in Eq. (\ref{67}) identity and the integral
\begin{equation}
\int d^3k\,\frac{4\,\pi\,\vec k}{k^4}\,
e^{i\,\vec k\cdot\vec r} = \frac{i}{2}\,\frac{\vec r}{r}\label{71}
\end{equation}
one obtains
\begin{eqnarray}
\delta H_B &=& \sum_{a>b}\sum_b\frac{\alpha}{4\,m^2}\biggl\{
\biggl[\vec\sigma_a\times\frac{\vec
    r_{ab}}{r_{ab}},\frac{p_a^2}{2\,m}\biggr]\,
    \bigl[V,\vec p_b] + \biggl[\frac{p_b^2}{2\,m},
\biggl[\vec\sigma_a\times\frac{\vec
    r_{ab}}{r_{ab}},\frac{p_a^2}{2\,m}\biggr]\biggr]\,\vec p_b 
\nonumber \\ &&
-\bigl[\vec p_a,V\bigr]\,\biggl[p_b^2,  \vec\sigma_b\times\frac{\vec
    r_{ab}}{r_{ab}}\biggr]- \vec p_a\,\biggl[\frac{p_a^2}{2\,m},
\biggl[\vec\sigma_a\times\frac{\vec
    r_{ab}}{r_{ab}}\,,\frac{p_b^2}{2\,m}\biggr]\biggr]\biggr\}\label{72}
\end{eqnarray}
Finally, the operator $\delta H_7$ is a sum of already derived parts
\begin{equation}
\delta H_7 = \delta H_A + \delta H_B + \delta H_C\label{73}
\end{equation}

$\delta H_8$ is a retardation correction in a single transverse
photon exchange, where one vertex is nonrelativistic, Eq. (\ref{49})
and the second comes from the 3rd term in Eq. (\ref{22})
\begin{equation}
-\frac{e}{8\,m^2}\,\vec\sigma\cdot\bigl(\vec E\times\vec p-\vec p\times\vec
 E\bigr)\label{74}
\end{equation}
With the help of Eq. (\ref{59}) one obtains the following expression for $\delta E_8$
\begin{eqnarray}
\delta E_8 &=& \sum_{a\neq b}\sum_b e^2\,\int\frac{d^3k}{(2\,\pi)^3}\,
\biggl(\delta^{ij}-\frac{k^i\,k^j}{k^2}\biggr)\,\frac{i}{16\,m^3}\,
\langle\phi|\Bigl(e^{i\,\vec k\cdot\vec r_a}\,\vec p_a\times\vec\sigma_a
+\vec p_a\times\vec\sigma_a\,e^{i\,\vec k\cdot\vec r_a}\Bigr)^i
\nonumber \\ &&\frac{1}{E_0-H_0-k}\,\biggl(\vec p_b -
\frac{i}{2}\,\vec\sigma_b\times\vec k\biggr)^j\,e^{-i\,\vec k\cdot\vec r_b}\,|\phi\rangle
+{\rm h.c.}\label{75}
\end{eqnarray}
In the expansion of $1/(E_0-H_0-k)$ in Eq. (\ref{60}) the first term vanishes
because h.c. and the second term is a correction of order $m\,\alpha^6$.
After commuting $(H_0-E_0)$ on the left one obtains the effective operator $\delta H_8$
\begin{eqnarray}
\delta H_8 &=& \sum_{a\neq b}\sum_b e^2\,\int\frac{d^3k}{(2\,\pi)^3}\,
\biggl(\delta^{ij}-\frac{k^i\,k^j}{k^2}\biggr)\,\frac{1}{k^2}\,\frac{i}{16\,m^3}
\nonumber \\ &&
\biggl[e^{i\,\vec k\cdot\vec r_{ab}}\,\vec p_a\times\vec\sigma_a
+\vec p_a\times\vec\sigma_a\,e^{i\,\vec k\cdot\vec r_{ab}},V+\frac{p_a^2}{2\,m}\biggr]^i
\,\biggl(\vec p_b - \frac{i}{2}\,\vec\sigma_b\times\vec k\biggr)^j
+{\rm h.c.}\nonumber \\
&=&\sum_a \frac{e^2}{8\,m^2}\,\vec\sigma_a\cdot
\Bigl[\vec{\cal E}_a\times \vec {\cal A}_{a} - 
\vec {\cal A}_{a}\times\vec{\cal E}_a\Bigr]
 \nonumber \\ &&
+\frac{i\,e}{16\,m^3}\,\Bigl[
 \vec {\cal A}_{a}\cdot\vec p_a\times\vec\sigma_a +
\vec p_a\times\vec\sigma_a\cdot \vec {\cal A}_{a}\,, p_a^2\Bigr]\label{76}
\end{eqnarray}

$\delta H_9$  is a one- and two-loop radiative correction
\begin{equation}
\delta H_9 = H_{R1} + H_{R2}\label{77}
\end{equation}
and its derivation requires a separate treatment. We base our treatment here on known
results for helium, which in turn are based on hydrogen and positronium, 
and extend it to an arbitrary atom, as long as nonrelativistic expansion makes sense.
\begin{eqnarray}
H_{R1} &=& \sum_a \frac{\alpha\,(Z\,\alpha)^2}{m^2}\,9.61837\,\delta^3(r_a)
\nonumber \\ &&
           +\sum_{a>b}\sum_b\frac{\alpha^3}{m^2}\,(14.33134 - 3.42651\,
           \vec\sigma_a\cdot\vec\sigma_b)\,\delta^3(r_{ab})\label{78}\\
H_{R2} &=& \sum_a \frac{\alpha^2\,(Z\,\alpha)}{\,m^2}\,0.17155\,\delta^3(r_a)
\nonumber \\ &&
           +\sum_{a>b}\sum_b\frac{\alpha^3}{m^2}\,(-0.66526 + 0.08633\,
           \vec\sigma_a\cdot\vec\sigma_b)\,\delta^3(r_{ab})
\nonumber \\ &&+\biggl(\frac{\alpha}{\pi}\biggr)^2\,\biggl\{\sum_a 
\frac{Z\,\alpha}{4\,m^2}\,
\vec\sigma_a\cdot\frac{\vec r_a}{r_a^3}\times \vec p_a\,2\,a_e^{(2)}
\nonumber \\ &&
+\sum_{a>b}\sum_b 
\frac{\alpha}{4\,m^2}\,\frac{\sigma_a^i\,\sigma_b^j}
{r_{ab}^3}\,
\biggl(\delta^{ij}-3\,\frac{r_{ab}^i\,r_{ab}^j}{r_{ab}^2}\biggr)\,
[2\,a_e^{(2)} + (a_e^{(1)})^2]
\nonumber \\ &&
-\frac{\alpha}{4\,m^2\,r_{ab}^3}\, 
\bigl(\vec\sigma_a+\vec \sigma_b\bigr)\,\cdot\vec r_{ab}\times(\vec p_a - \vec
p_b)\,2\,a_e^{(2)}\biggr\}\,.\label{79}
\end{eqnarray}
where
\begin{eqnarray}
\kappa    & = & \frac{\alpha}{\pi}\,a_e^{(1)} +
\biggl(\frac{\alpha}{\pi}\biggr)^2\,a_e^{(2)}
+ \ldots \label{80}\\
a_e^{(1)} & = & 0.5\,,\label{81}\\
a_e^{(2)} & = & -0.328\,478\,965\ldots\label{82}
\end{eqnarray}
and $\kappa$ is the electron magnetic moment anomaly. 

\hspace{5in}
\section{Summary} 
The obtained complete $m\,\alpha^6$ contribution $H^{(6)}$ Eq. (\ref{38})
is in agreement with the former derivation for the particular case
of helium $S$ \cite{helium1} and $P$ levels \cite{dk}, but is much more compact.
Due to differing ways of representing various complicated operators,
this comparison is rather nontrivial, and we had to refer
in many case to momentum representation to find agreement.
Since the present derivation differs from the former one,
this agreement may be regarded as a justification of
this, and as well as the former results. 
We have not derived here the term $\lambda$ in Eq. (\ref{05}).
It is obtained by matching the forward scattering amplitude
in full QED with the one obtained from the effective Hamiltonian or NRQED, 
and it accounts for the contribution with high electron momentum.
$\lambda$ depends however, on the regularization scheme, and once it is fixed
it can be obtained in a similar way as in dimensional \cite{ky} or 
photon propagator regularizations \cite{helium1}.

$H^{(6)}$ can be used for high precision
calculations of energy levels of few electron atoms, provided
two difficulties are overcome. The first one is the algebraic elimination
of electron-electron singularities. The elimination of electron-nucleus
singularities was demonstrated on hydrogen and helium examples,
and could easily be extended to an arbitrary atom. The elimination
of electron-electron singularities using the dimensional regularization scheme
was performed for the ground state of helium atom in \cite{ky}, however
this derivation was very complicated and so far this result has not been 
confirmed. The extension to more than two-electron atoms is even more
complicated, therefore a new idea which will lead to 
elimination of electron-electron singularities is needed.
The second difficulty is the lack of analytical values for integrals
with basis sets, which fulfill the cusp conditions. For example,
for the Hylleraas basis set only three-electron integrals are known analytically
\cite{remiddi, krprem}. 
This cusp condition is necessary, because effective operators present
in $\delta H^{(6)}$ contain many derivatives. For these reasons the calculation
of $m\,\alpha^6$ contribution to atomic energy levels has been accomplished
only for a few states of helium atoms \cite{yandrake,slamb,ky,plamb} 
and hyperfine splitting \cite{2shfs}, where a powerful random
exponential basis set has been applied. For the three- \cite{lithium1,lithium2} 
and four-electron \cite{berylium}
systems the leading QED effects,
namely the correction of order $m\,\alpha^5$ have only recently been
calculated. Due to developments
in the Hylleraas basis sets \cite{yan, krprem} we think the calculation
of $m\,\alpha^6$ contribution to lithium energy levels is now possible, particularly
interesting is the $Q(\alpha^2)$ correction to hyperfine splitting, which 
may be regarded as a benchmark calculations for MCDF or MBPT methods.
Another interesting example is the fine structure of $P_J$ levels, 
where electron-electron singularities are not present, such as for
helium fine structure.
  
\section{Acknowledgments}
I wish to thank ENS Paris for supporting my visit in 
Laboratoire Kastler-Brossel, where this work has been written,
and I acknowledge interesting discussions with Paul Indelicato
and Jan Derezi\'nski.
This was was supported in part by  Postdoctoral Training Program HPRN-CT-2002-0277.

\end{document}